\documentclass[12pt]{article}
\pdfoutput=1
\usepackage{makeidx}
\makeindex
\usepackage[a4paper]{geometry}
\usepackage{jheppub,amsmath,amssymb,amsfonts,amsxtra, mathrsfs,graphics,graphicx,amsthm,epsfig,ytableau,bm,longtable,float,color,tikz,mathtools,xfrac,footnote,rotating,lscape}
\usepackage{textgreek}
\usetikzlibrary{decorations.pathmorphing}
\usetikzlibrary{decorations.markings}
\usetikzlibrary{arrows, decorations.markings, calc, fadings, decorations.pathreplacing, patterns, decorations.pathmorphing, positioning}
\usepackage{tikz-cd}

\usetikzlibrary{positioning,shapes}
\usetikzlibrary{chains}
\usetikzlibrary{arrows,fit,decorations.pathreplacing}
\tikzstyle{every picture}+=[remember picture]
\tikzstyle{na} = [baseline=-.5ex]

\usepackage{empheq}
\usepackage{multirow}
\usepackage{booktabs}
\usepackage[american]{babel}

\usepackage[latin1]{inputenc}

\makeatletter
\newcommand{\vast}{\bBigg@{1}}
\newcommand{\Vast}{\bBigg@{5}}
\makeatother

\usepackage{hyperref}

\numberwithin{equation}{section}

\newcommand{\ie}{\textit{i.e.}}

\numberwithin{equation}{section}

\newcommand{\be}{\begin{equation}} \newcommand{\ee}{\end{equation}}
\newcommand{\bea}{\begin{equation} \begin{aligned}} \newcommand{\eea}{\end{aligned} \end{equation}}

\def\SO{\mathrm{SO}}

\def\U{\mathrm{U}}
\def\SU{\mathrm{SU}}

\newcommand{\rd}{\mathrm{d}}

\DeclareMathOperator{\Tr}{Tr}

\DeclareMathOperator{\re}{\mathbb{R}e}
\DeclareMathOperator{\im}{\mathbb{I}m}

\newcommand{\cF}{\mathcal{F}}
\newcommand{\cG}{\mathcal{G}}
\newcommand{\cH}{\mathcal{H}}
\newcommand{\cI}{\mathcal{I}}
\newcommand{\cJ}{\mathcal{J}}
\newcommand{\cK}{\mathcal{K}}
\newcommand{\cL}{\mathcal{L}}
\newcommand{\cM}{\mathcal{M}}
\newcommand{\cN}{\mathcal{N}}

\newcommand{\cS}{\mathcal{S}}

\newcommand{\cV}{\mathcal{V}}

\newcommand{\cX}{\mathcal{X}}

\newcommand{\cZ}{\mathcal{Z}}

\newcommand{\bC}{\mathbb{C}}

\newcommand{\bP}{\mathbb{P}}

\newcommand{\bR}{\mathbb{R}}

\usepackage[Symbol]{upgreek}
\usepackage{bm}

\usepackage{calligra}
\DeclareMathAlphabet{\mathcalligra}{T1}{calligra}{m}{n}

\title{
An extremization principle for the entropy of rotating BPS black holes in AdS$_5$}

\author[a,b]{Seyed Morteza Hosseini,}
\author[c]{Kiril Hristov}
\author[a,b]{and Alberto Zaffaroni}

\affiliation[a]{Dipartimento di Fisica, Universit\`a di Milano-Bicocca, I-20126 Milano, Italy}
\affiliation[b]{INFN, sezione di Milano-Bicocca, I-20126 Milano, Italy}
\affiliation[c]{Institute for Nuclear Research and Nuclear Energy, Bulgarian Academy of Sciences, \\Tsarigradsko Chaussee 72, 1784 Sofia, Bulgaria}

\emailAdd{morteza.hosseini@mib.infn.it}
\emailAdd{khristov@inrne.bas.bg}
\emailAdd{alberto.zaffaroni@mib.infn.it}

\abstract{We show that the Bekenstein-Hawking entropy of a class of  BPS electrically charged  rotating black holes in AdS$_5\times S^5$ can be obtained by a simple extremization principle. We expect that this extremization corresponds to the attractor mechanism for BPS rotating black holes in five-dimensional gauged supergravity, which is still unknown. The expression to be extremized has a suggestive resemblance to anomaly polynomials and the supersymmetric Casimir energy recently studied for  ${\cal N}=4$ super Yang-Mills.
}

\begin{document}

\setcounter{tocdepth}{2}
\maketitle

%
%

\date{Dated: \today}




\section{Introduction}
\label{sec:introduction}

The derivation of the entropy of BPS electrically charged  rotating black holes in
AdS$_5\times S^5$ \cite{Gutowski:2004ez,Gutowski:2004yv,Chong:2005da,Chong:2005hr,Kunduri:2006ek}
in terms of states of the dual ${\cal N}=4$ $\SU(N)$ super Yang-Mills (SYM) theory is still an open problem.
Various attempts have been made in this direction \cite{Kinney:2005ej,Grant:2008sk,Chang:2013fba} but none was really successful.
One could consider the superconformal index \cite{Romelsberger:2005eg,Kinney:2005ej} since it counts states preserving
the same supersymmetries of the black holes and it depends on a number of fugacities equal to the number of conserved charges of the black holes.
However, due to a large cancellation between bosonic and fermionic states, the index is a quantity of order one
for generic values of the fugacities while the entropy scales like $N^2$ \cite{Kinney:2005ej}.
We also know that the supersymmetric partition function on $S^3\times S^1$ is equal to the superconformal index
only up to a multiplicative factor $e^{-\beta E_{\rm SUSY}}$, where the supersymmetric Casimir energy $E_{\rm SUSY}$
is of order $N^2$ \cite{Assel:2014paa,Lorenzen:2014pna,Assel:2015nca,Bobev:2015kza,Martelli:2015kuk,Genolini:2016sxe,Brunner:2016nyk}.
However, it is not clear what the average energy of the vacuum should have to do with the entropy, which is the degeneracy of ground states.
The analogous problem for static, asymptotically AdS$_4$ dyonic black holes was recently solved in \cite{Benini:2015eyy,Benini:2016rke}.
It was shown in \cite{Benini:2015eyy,Benini:2016rke} that the topologically twisted index for three-dimensional
gauge theories \cite{Benini:2015noa}\footnote{For further developments
see \cite{Benini:2016hjo,Cabo-Bizet:2016ars,Closset:2016arn,Closset:2017zgf}.}
has no large cancellation between bosons and fermions at large $N$,
it scales like $N^{3/2}$ \cite{Benini:2015eyy,Hosseini:2016tor,Hosseini:2016ume} (see also \cite{Cabo-Bizet:2017jsl})
and correctly reproduces the entropy of a class of BPS black holes in AdS$_4\times S^7$.
More precisely, the topologically twisted index $Z_{\rm twisted}$ is a function of magnetic charges $p_i$ and
fugacities $\Delta_i$  for the global symmetries of the theory.
The entropy of the black holes with electric charges $q_i$ is
then obtained as a Legendre transform of $\log Z_{\rm twisted}$:
\be
 \label{Legendre}
 S(q_i,p_i) = \log Z_{\rm twisted}(p_i,\Delta_i) - i \sum_i q_i \Delta_i \Big |_{\bar\Delta_i} \, ,
\ee  
where $\bar\Delta_i$ is the extremum of ${\cal I}(\Delta_i)=\log Z_{\rm twisted}(p_i,\Delta_i)  - i q_i \Delta_i$. This procedure has been called ${\cal I}-$extremization in \cite{Benini:2015eyy,Benini:2016rke} and shown to correspond to the attractor mechanism in gauged supergravity \cite{Ferrara:1996dd,Cacciatori:2009iz,DallAgata:2010ejj,Hristov:2010ri,Katmadas:2014faa,Halmagyi:2014qza,Klemm:2016wng}. 

It is natural to ask what would be the analogous of this construction in five dimensions.
In this paper, we humbly look attentively at the gravity side of the story and try to understand
what kind of extremization can reproduce the entropy of the supersymmetric rotating black holes.
Unfortunately, the details of the attractor mechanism for rotating black holes in five-dimensional
gauged supergravity are not known but we can nevertheless find an extremization principle for the entropy.
The final result is quite  surprising and intriguing. 

We consider  the class of supersymmetric rotating black holes found and studied in \cite{Gutowski:2004ez,Gutowski:2004yv,Chong:2005da,Chong:2005hr,Kunduri:2006ek}. They are asymptotic to  AdS$_5\times S^5$
and depend on three electric charges  $Q_I$ $(I = 1,2,3)$, associated with rotations in $S^5$, and two angular momenta $J_\phi, J_\psi$ in AdS$_5$.
Supersymmetry actually requires a constraint among the charges and only four of them are independent.
We show that the Bekenstein-Hawking entropy of the black holes can be obtained as the Legendre transform
of the quantity\footnote{Notice that one can write the very same entropy as the result of a different extremization
in the context of the Sen's entropy functional \cite{Sen:2005wa,Morales:2006gm}.
The two extremizations are over different quantities and use different charges.}  
\bea
  \label{modified:Casimir:SUSY:N=4}
 E= - i \pi  N^2  \frac{\Delta_1 \Delta_2 \Delta_3}{\omega_1 \omega_2} \, ,
\eea
where $\Delta_I$ are  chemical potentials conjugated to the electric charges $Q_I$ and $\omega_{1,2}$
chemical potentials conjugated to the angular momenta $J_\phi, J_\psi$.
The constraint among charges is reflected in the following constraint among chemical potentials,
\bea
 \label{constraint:N=4}
 \Delta_1 + \Delta_2 + \Delta_3 + \omega_1 + \omega_2 = 1 \, .
\eea
To further motivate the result  \eqref{modified:Casimir:SUSY:N=4} we shall consider the case of equal angular momenta $J_\psi=J_\phi$.
In this limit, the black hole has an enhanced $\SU(2)\times \U(1)$ isometry
and it can be reduced along the $\U(1)$ to a static dyonic black hole in four dimensions.
We show that, upon dimensional reduction, the extremization problem
based on \eqref{modified:Casimir:SUSY:N=4} coincides with the attractor mechanism in four dimensions,
which is well understood for static BPS black holes \cite{Ferrara:1996dd,Cacciatori:2009iz,DallAgata:2010ejj,Hristov:2010ri,Katmadas:2014faa,Halmagyi:2014qza,Klemm:2016wng}. 

It is curious to observe that the expression \eqref{modified:Casimir:SUSY:N=4} is {\it formally} identical
to the supersymmetric Casimir energy for ${\cal N}=4$ SYM, as derived, for example, in \cite{Bobev:2015kza} and reviewed in appendix \ref{AppC}.
It appears in the relation $Z_{{\cal N}=4}=e^{- E} I$ between the partition function $Z_{{\cal N}=4}$ on $S^3\times S^1$
and the superconformal index   $I$. Both the partition function and the superconformal index are functions
of a set of chemical potentials $\Delta_I$ $(I=1,2,3)$ and $\omega_i$ $(i = 1,2)$ associated with the R-symmetry generators $\U(1)^3\in \SO(6)$ and
the two angular momenta $\U(1)^2\in \SO(4)$, respectively. Since the symmetries that appear in the game must commute with
the preserved supercharge, the index and the partition function are actually functions of only four independent chemical potentials,
precisely as our quantity $E$. The constraint among chemical potentials is usually imposed as $\sum_{I=1}^3 \Delta_I +\sum_{i=1}^2 \omega_i =0$.
Since chemical potentials in our notations are periodic of period 1, our constraint \eqref{constraint:N=4} reflects a different
choice for the angular ambiguities. We comment about  the interpretation of this result in the discussion section,
leaving the proper understanding to future work.
 
The paper is  organized as follows. In section \ref{sec:AdS5 black holes}
we review the basic features of the BPS rotating black holes of interest.
In section \ref{sec:Casimir and entropy}, we show that the Bekenstein-Hawking entropy of
the black hole can be obtained as the Legendre transform of the quantity \eqref{modified:Casimir:SUSY:N=4}.
In section  \ref{sec:a limiting case}, we perform the dimensional reduction of
the black holes with equal angular momenta down to four dimensions and we prove that
the extremization of \eqref{modified:Casimir:SUSY:N=4} is equivalent to
the attractor mechanism for four-dimensional static BPS black holes in gauged supergravity.
We conclude in section \ref{sec:discussion} with discussions and future directions. 
The appendices contain the relevant information and conventions about gauged supergravity
in five and four dimensions and the supersymmetric Casimir energy.

\section{Supersymmetric AdS$_5$ black holes in $\U(1)^3$ gauged supergravity}
\label{sec:AdS5 black holes}

In this section we will briefly review a class of supersymmetric, asymptotically AdS, black holes
of $D = 5$ $\U(1)^3$ gauged supergravity \cite{Gutowski:2004ez,Gutowski:2004yv,Chong:2005da,Chong:2005hr,Kunduri:2006ek}.
They can be embedded in type IIB supergravity as an asymptotically AdS$_5 \times S^5$ solution
which is exactly the decoupling limit of the rotating D3-brane \cite{Cvetic:1999xp}.
Further details about the supergravity model can be found in \cite{Cvetic:1999xp} and appendix \ref{app:5d gauged sugra}.
When lifted to type IIB supergravity they preserve only two real supercharges \cite{Gauntlett:2004cm}.
They are characterized by their mass, three electric charges and two  angular momenta with a constraint, and are
holographically dual to $1/16$ BPS states of $\cN = 4$ $\SU(N)$ SYM theory on $S^3 \times \mathbb{R}$ at large $N$.

We shall primarily be interested in the so-called $\cN = 2$ gauged STU model,
that arises upon compactification of type IIB supergravity on $S^5$.
In the notations of appendix \ref{app:5d gauged sugra}, the only nonvanishing triple intersection numbers are $C_{1 2 3} = 1$ (and cyclic permutations).
The bosonic sector of the theory comprises three gauge fields which correspond to the Cartan subalgebra of the $\SO(6)$ isometry of $S^5$,
the metric, and three real scalar fields subject to the constraint
\be
 \label{XI:constraint}
 L^1 L^2 L^3 = 1 \, .
\ee
They take vacuum values $\bar L^I = 1$. 
The five-dimensional black hole metric can be written as \cite{Kunduri:2006ek}
\be
 \label{AdS5:generic:metric}
 ds^2 = - (H_1 H_2 H_3)^{-2/3} \left( dt + \omega_\psi d \psi + \omega_\phi d \phi \right)^2 + \left( H_1 H_2 H_3 \right)^{1/3} h_{m n} dx^m dx^n \, ,
\ee
where
\be
 H_I = 1 + \frac{\sqrt{\Xi_a \Xi_b} \left( 1 + g^2 \mu_I \right) - \Xi_a \cos^2 \theta - \Xi_b \sin^2 \theta }{g^2 r^2} \, ,
\ee
\bea
 h_{m n} dx^m dx^n & = r^2 \bigg\{ \frac{dr^2}{\Delta_r} + \frac{d\theta^2}{\Delta_\theta}
 + \frac{\cos^2\theta}{\Xi_b^2} \left[ \Xi_b + \cos^2\theta \left( \rho^2 g^2 + 2 \left( 1+ b g \right) \left( a + b \right) g \right) \right] d\psi^2 \\
 & + \frac{\sin^2\theta}{\Xi_a^2} \left[ \Xi_a + \sin^2\theta \left( \rho^2 g^2 + 2 \left( 1+ a g \right) \left( a + b \right) g \right) \right] d\phi^2 \\
 & + \frac{2 \sin^2\theta \cos^2\theta}{\Xi_a \Xi_b} \left[ \rho^2 g^2 + 2 \left( a + b \right) g + \left( a + b \right)^2 g^2 \right] d\psi d\phi \bigg\} \, ,
\eea
\bea
 \Delta_r & = r^2 \left[ g^2 r^2 + \left( 1 + a g + b g \right)^2 \right] \, , \qquad \qquad \; \;
 \Delta_\theta = \Xi_a \cos^2\theta + \Xi_b \sin^2\theta \, , \\
 \Xi_a & = 1 - a^2 g^2 \, , \qquad \Xi_b = 1 - b^2 g^2 \, , \qquad \qquad \rho^2 = r^2 + a^2 \cos^2\theta + b^2 \sin^2\theta \, ,
\eea
\bea
 \omega_\psi & = - \frac{g \cos^2\theta}{r^2 \Xi_b} \left[ \rho^4 + \left( 2 r_m^2 + b^2 \right) \rho^2 + \frac12 \left( \beta_2 - a^2 b^2 + g^{-2} \left( a^2 - b^2 \right) \right) \right] \, , \\
 \omega_\phi & = - \frac{g \sin^2\theta}{r^2 \Xi_a} \left[ \rho^4 + \left( 2 r_m^2 + a^2 \right) \rho^2 + \frac12 \left( \beta_2 - a^2 b^2 - g^{-2} \left( a^2 - b^2 \right) \right) \right] \, ,
\eea
and
\bea
 r_m^2 & = g^{-1} (a +b) + a b \, , \\
 \beta_2 & = \Xi_a \Xi_b \left( \mu_1 \mu_2 + \mu_1 \mu_3 + \mu_2 \mu_3 \right)
 - \frac{2 \sqrt{\Xi_a \Xi_b} \left(1 - \sqrt{\Xi_a \Xi_b} \right)}{g^2} \left( \mu_1 + \mu_2 + \mu_3 \right) + \frac{3 \left(1 - \sqrt{\Xi_a \Xi_b} \right)^2}{g^4} \, .
\eea
The gauge coupling constant $g$ is fixed in terms of the AdS$_5$ radius of curvature, $g = 1 / \ell$.
The coordinates are $\left( t, r, \theta, \phi, \psi \right)$ where $r > 0$ corresponds to the exterior of the black hole, $0 \leq \theta \leq \pi / 2$ and $0 \leq \phi, \psi \leq 2 \pi$.
The scalars read
\be
 L^I = \frac{\left( H_1 H_2 H_3 \right)^{1/3}}{H_I} \, ,
\ee
while the gauge potentials are given by
\be
 A^I = H_I^{-1} \left( dt + \omega_\psi d\psi + \omega_\phi d \phi \right) + U_\psi^I d\psi + U_\phi^I d\phi \, ,
\ee
where
\bea
 U^I_\psi & = \frac{g \cos^2\theta}{\Xi_b} \left[ \rho^2 + 2 r_m^2 + b^2 - \sqrt{\Xi_a \Xi_b} \mu_I + g^{-2} \left( 1 - \sqrt{\Xi_a \Xi_b} \right) \right] \, , \\
 U_\phi^I & = \frac{g \sin^2\theta}{\Xi_a} \left[ \rho^2 + 2 r_m^2 + a^2 - \sqrt{\Xi_a \Xi_b} \mu_I + g^{-2} \left( 1 - \sqrt{\Xi_a \Xi_b} \right) \right] \, .
\eea
The black hole is labeled by five parameters: $\mu_{1,2,3}, a, b$ where $g^{-1} > a$, $b \geq 0$.
Only four parameters are independent due to the constraint
\be
 \label{gravity:constraint}
 \mu_1 + \mu_2 + \mu_3 = \frac{1}{\sqrt{\Xi_a \Xi_b}} \left[ 2 r_m^2 + 3 g^{-2} \left( 1 - \sqrt{\Xi_a \Xi_b} \right) \right] \, .
\ee
Furthermore, regularity of the scalars for $r \geq 0$ entails that
\be
 g^2 \mu_I > \sqrt{\frac{\Xi_b}{\Xi_a}} - 1 \geq 0 \, ,
\ee
when $a \geq b$. If $a < b$ then the same expression remains valid with $a$ and $b$ interchanged.

\subsection{The asymptotic AdS$_5$ vacuum}

This solution is expressed in the co-rotating frame.
The change of coordinates  $t = \bar t$, $\phi = \bar\phi - g \bar t$, $\psi = \bar\psi - g \bar t$, and $y^2 = r^2 + 2 r_m^2 / 3$
transforms the metric to a static frame at infinity. In order to bring the metric into
a manifestly asymptotically AdS$_5$ spacetime (in the global sense) as $y \to \infty$
we make the following change of coordinates \cite{Kunduri:2006ek}
\bea
  \Xi_a Y^2 \sin^2\Theta = \left( y^2 + a^2 \right) \sin^2 \theta \, , \qquad \qquad \Xi_b Y^2 \cos^2\Theta = \left( y^2 + b^2 \right) \cos^2 \theta \, .
\eea
One gets the line element
\bea
 ds^2 \simeq - g^2 Y^2 \, d\bar{t}^2 + \frac{dY^2}{g^2 Y^2} + Y^2 \left( d\Theta^2 + \sin^2\Theta \, d\bar\phi^2 + \cos^2\Theta \, d\bar\psi^2 \right) \, .
\eea
The black hole has the Einstein universe $\bR \times S^3$ as its conformal boundary.
In the asymptotically static coordinates, the supersymmetric Killing vector field reads
\bea
 V = \frac{\partial}{\partial \bar{t}} + g \frac{\partial}{\partial\bar{\phi}} + g \frac{\partial}{\partial\bar{\psi}} \, ,
\eea
which is timelike everywhere outside the black hole and is null on the conformal boundary.

\subsection{Properties of the solution}

It is convenient to define the following polynomials,
\be
 \gamma_1 = \mu_1 + \mu_2 + \mu_3 \, , \qquad \gamma_2 = \mu_1 \mu_2 + \mu_1 \mu_3 + \mu_2 \mu_3 \, , \qquad \gamma_3 = \mu_1 \mu_2 \mu_3 \, .
\ee
The black hole carries three $\U(1)^3 \subset \SO(6)$ electric charges in $S^5$ which are given by
\bea
 \label{BH:Q:5d:U(1)^3}
 Q_I = \frac{\pi}{4 G_{\rm N}^{(5)}} \left[ \mu_I + \frac{g^2}{2} \left( \gamma_2 - \frac{2 \gamma_3}{\mu_I} \right) \right] \, , \quad
 \text{ for } \quad I = 1, 2, 3 \, ,
\eea
and two $\U(1)^2 \subset \SO(4)$ angular momenta in AdS$_5$ that read
\bea
 \label{BH:J:5d:U(1)^3}
 J_\psi & = \frac{\pi}{4 G_{\rm N}^{(5)}} \left[ \frac{g \gamma_2}{2} + g^3 \gamma_3 + g^{-3} \left(\sqrt{\frac{\Xi_a}{\Xi_b}} - 1 \right) \cJ \right] \, , \\
 J_\phi & = \frac{\pi}{4 G_{\rm N}^{(5)}} \left[ \frac{g \gamma_2}{2} + g^3 \gamma_3 + g^{-3} \left(\sqrt{\frac{\Xi_b}{\Xi_a}} - 1 \right) \cJ \right] \, .
\eea
Here, $G_{\rm N}^{(5)}$ is the five-dimensional Newton constant and we defined
\be
 \cJ \equiv \prod_{I = 1}^{3} \left( 1 + g^2 \mu_I \right) \, .
\ee
The mass of the black holes is determined by the BPS condition
\bea
 \label{BH:BPS relation}
 M = g |J_\phi| + g |J_\psi| + |Q_1| + |Q_2| + |Q_3| \, ,
\eea
which yields
\bea
 \label{BH:M:5d:U(1)^3}
 M = \frac{\pi}{4 G_{\rm N}^{(5)}} \left[
 \gamma_1 + \frac{3 g^2 \gamma_2}{2} + 2 g^4 \gamma_3
 + \frac{\left( \sqrt{\Xi_a} - \sqrt{\Xi_b} \right)^2}{g^2 \sqrt{\Xi_a \Xi_b}} \cJ \right] \, . ~~
\eea
The solution has a regular event horizon at $r_{\rm h} = 0$ only for nonzero angular momenta in AdS$_5$.
The angular velocities of the horizon, measured with respect to the azimuthal coordinates $\psi$ and $\phi$
of the asymptotically static frame at infinity, are
\be
 \Omega_\psi = \Omega_\phi = g \, .
\ee
The near-horizon geometry is a fibration of AdS$_2$ over a non-homogeneously squashed $S^3$ \cite{Kunduri:2007qy} with area
\bea
 \label{BH:area:5d:U(1)^3}
 \text{Area} = 2 \pi^2
 \sqrt{ \gamma_3 \left( 1 + g^2 \gamma_1 \right) - \frac{g^2 \gamma_2^2}{4}
 - \frac{\left( \sqrt{\Xi_a} - \sqrt{\Xi_b} \right)^2}{g^6 \sqrt{\Xi_a \Xi_b}} \cJ} \, . ~~
\eea
Positivity of the expression within the square root ensures the absence of closed causal curves near $r = 0$.
The Bekenstein-Hawking entropy of the black hole is proportional to its horizon area
and can be compactly written in terms of the physical charges as \cite{Kim:2006he}
\bea
 \label{BH:entropy:5d:U(1)^3}
 S_{\text{BH}} & = \frac{\text{Area}}{4 G_{\rm N}^{(5)}}
 = \frac{2 \pi}{g} \sqrt{Q_1 Q_2 + Q_2 Q_3 + Q_1 Q_3 - \frac{\pi}{4 G_{\rm N}^{(5)} g} \left( J_\phi + J_\psi \right)} \, .
\eea
Finally, let
\bea
 \cX_I = \left( 1 + g^2 \mu_I \right) \sqrt{\Xi_a \Xi_b} - \Delta_\theta \, .
\eea
The values of the scalar fields at the horizon read
\bea
 L^I (r_{\rm h}) = \frac{\left( \cX_1 \cX_2 \cX_3 \right)^{1/3}}{\cX_I} \, .
\eea

In the next section we will obtain  the Bekenstein-Hawking entropy \eqref{BH:entropy:5d:U(1)^3} of the BPS black hole
from an extremization principle.

\section{An extremization principle for the  entropy}
\label{sec:Casimir and entropy}

We shall now extremize the quantity  \eqref{modified:Casimir:SUSY:N=4}, 
and show that the extremum precisely reproduces the entropy of the multi-charge BPS black hole discussed in the previous section.

Let us first introduce some notation that facilitates the comparison with supergravity:
\bea
 X^I \equiv \Delta_I \, , \qquad \qquad X^0_{\pm} \equiv \omega_1 \pm \omega_2 \, ,
\eea
where $I = 1, 2 , 3$.
We shall also use $J^{\pm} \equiv J_\phi \pm J_\psi$,
\bea
 J^{+} & = \frac{\pi}{4 G_{\text{N}}^{(5)}} \left[ g \gamma_2 + 2 g^3 \gamma_3 + \frac{\left( \sqrt{\Xi_a} - \sqrt{\Xi_b} \right)^2}{g^3 \sqrt{\Xi_a \Xi_b}} \cJ \right] \, , \\
 J^{-} & = \frac{\pi}{4 G_{\text{N}}^{(5)}} \frac{\Xi_b - \Xi_a}{g^3 \sqrt{\Xi_a \Xi_b}} \cJ \, .
\eea
Thus, we can rewrite the quantity  \eqref{modified:Casimir:SUSY:N=4}  as
\bea
\label{eq:Casimir:rewrittenX}
 E= - \frac{2 \pi^2 i}{g^3 G_{{\rm N}}^{(5)}} \frac{X^1 X^2 X^3}{\left( X^0_{+} \right)^2 - \left( X^0_{-} \right)^2} \, ,
\eea
where we used the standard relation between gravitational and QFT parameters in the large $N$ limit,
\bea
 \frac{\pi}{2 g^3 G_{{\rm N}}^{(5)}} = N^2 \, .
\eea
In the following we set the unit of the AdS$_5$ curvature $g = 1$.
The entropy of the BPS black hole, at leading order, can be obtained by extremizing the quantity\footnote{This is not the only possible choice of signs. There
are various sign ambiguities in the superconformal index literature as well as in the black hole one that should be fixed in a proper comparison between gravity and field theory.}
\bea
 \label{rotating:attractor:4d}
 \cS = - E\left( X_{\pm}^0, X^I \right) + 2 \pi i \sum_{I = 1}^{3} Q_I X^I - \pi i \left( J^{+} X_{+}^0 + J^{-} X^0_{-} \right) 
  \, ,
\eea
with respect to $X^I$, $X_{\pm}^0$ and subject to the constraint \eqref{constraint:N=4},
\bea
\label{eq:constraintX}
 X_{+}^0 + \sum_{I = 1}^{3} X^I = 1 \, .
\eea
At this stage, we find it more convenient to work in the basis $z^{\alpha}$ $(\alpha = 0 , 1 , 2 , 3)$ which is related to $\left( X_{\pm}^0 , X^I \right)$ by
\bea
 X_{-}^{0} & = \frac{z^0}{1 + z^1 + z^2 + z^3} \, , \qquad \qquad X_{+}^{0} = \frac{1}{1 + z^1 + z^2 + z^3} \, , \\
 X^{1,2,3} & = \frac{z^{1,2,3}}{1 + z^1 + z^2 + z^3} \, .
\eea
Hence, in terms of the variables $z^\alpha$ the extremization equations can be written as
\bea
 \label{extremization:scalars}
 & \left[ (z^0)^2 -1 \right] \left\{ \left[ (z^0)^2 - 1 \right] c^i + \frac{z^1 z^2 z^3}{z^i} \right\} - 2 z^1 z^2 z^3 = 0 \, , \quad \text{ for } \quad i = 1, 2, 3 \, ,\\
 & c^0 \left[ (z^0)^2 - 1 \right]^2 - 2 z^0 z^1 z^2 z^3 = 0 \, ,
\eea
where we defined the constants
\bea
 c^0 = \frac{\cJ \left( \Xi_b - \Xi_a \right)}{8 \sqrt{\Xi_a \Xi_b}} \, , \qquad \qquad
 c^i = \frac{\cJ}{4} \left( \frac{1}{1 + \mu_i} - \frac{\Xi_b + \Xi_a}{2 \sqrt{\Xi_a \Xi_b}} \right) \, .
\eea
With an explicit computation one can check that the value of $\cS\left( z^\alpha \right)$
at the critical point precisely reproduces the entropy of the black hole,
\bea
\label{eq:EntropyMatch}
  \cS \big|_{\text{crit}} \left( J^{\pm} , Q_I \right) = S_\text{BH} \left( J^{\pm} , Q_I \right) \, .
 \eea
It is remarkable that the solution to the extremization equations \eqref{extremization:scalars} is complex; however,
at the saddle-point it becomes a real function of the black hole charges.
We conclude that  the extremization of the quantity  \eqref{modified:Casimir:SUSY:N=4}  yields exactly the
Bekenstein-Hawking entropy of the $1 / 16$ BPS black holes in AdS$_5 \times S^5$.

So far the discussion was completely general.
In the next section, we will analyze the case $J_\phi = J_\psi$, for which the solution to
the extremization equations takes a remarkably simple form.

\section{Dimensional reduction in the limiting case: $J_\phi = J_\psi$}
\label{sec:a limiting case}

We gain some important insight by considering the dimensional reduction of the
five-dimensional BPS black holes when the two angular momenta are equal.
The black hole metric on the squashed sphere then has an enhanced isometry $\SU(2) \times \U(1) \subset \SO(4)$.
If we choose the appropriate Hopf coordinates we can dimensionally reduce the solution along the $\U(1)$
down to four-dimensional gauged supergravity.
As discussed in \cite{Hristov:2014eza}, it turns out that such a dimensional
reduction makes sense not only for asymptotically flat solutions where
first discovered in \cite{Gaiotto:2005gf,Gaiotto:2005xt,Behrndt:2005he,Banerjee:2011ts}
but also for the asymptotically AdS solutions in the gauged supergravity considered here.
A crucial difference is that the lower-dimensional vacuum will no longer be maximally symmetric
but will instead be of the hyperscaling-violating Lifshitz (hvLif) type \cite{Hristov:2014eza}.

The reason for looking at  the limit $J_\phi = J_\psi$ is simple: due to the $\SU(2)$ symmetry  the lower-dimensional solution
is guaranteed to be static and the horizon metric is a direct product AdS$_2 \times S^2$ geometry, as will be shown in due course.
Since the attractor mechanism for static BPS black holes in four-dimensional $\cN = 2$ gauged supergravity has been completely
understood \cite{Cacciatori:2009iz,DallAgata:2010ejj,Hristov:2010ri} we can fit the reduced solution in this framework.

\subsection{The near-horizon geometry}
\label{ssec:nh geometry}

We begin by taking the near-horizon limit, $r \to 0$, of the BPS black hole solution presented in section \ref{sec:AdS5 black holes}.
We set $a = b$, corresponding to the equal angular momenta $(J_{\psi} = J_{\phi})$,
and adopt the notation $\Xi_a = \Xi_b \equiv \Xi$.

Let us first introduce the following coordinates,
\be
 \psi \equiv \frac12 (\chi + \varphi) \, , \qquad \qquad \phi \equiv \frac12 (\chi - \varphi) \, , \qquad \qquad \theta \equiv \frac12 \vartheta \, ,
\ee
where $\vartheta, \varphi, \chi$ are the Euler angles of $S^3$ with $0 \leq \vartheta \leq \pi$, $0 \leq \varphi < 2 \pi$, $0 \leq \chi < 4\pi$.
The near-horizon geometry then reads
\bea
 \label{J1:J2:near-horizon:AdS2:M3}
 ds^2 & = R_{\text{AdS}_2}^2 ds^2_{\text{AdS}_2}
 + \gamma_3^{1/3} \, ds^2_{\cM_3} \, , \\
 L^I & = \frac{\gamma_3^{1/3}}{\mu_I} \, , \qquad \qquad
 A^I = \frac{\gamma_3^{1/3}}{\mu_I} R_{\text{AdS}_2} \, \tilde r \, d \tilde{t}
 + g \left( \gamma_1 - \mu_I - \frac{\gamma_2}{2 \mu_I} \right) \, d\gamma \, ,
\eea
where we defined
\bea
 ds^2_{\text{AdS}_2} & = - \tilde r^2 d\tilde t^2 + \frac{d\tilde r^2}{\tilde r^2} \, , \qquad \qquad
 R_{\text{AdS}_2}^2 = \frac{\gamma_3^{1/3} }{4 (1 + g^2 \gamma_1 )} \, , \\
 \tilde r & = \frac{r^2}{4 R_{\text{AdS}_2}^2} \, , \qquad \qquad \qquad
 \tilde t = \frac{2}{\Xi \sqrt{\gamma_3^{1/3} (1 + g^2 \gamma_1) }} \, t \, ,
\eea
\bea
 \label{J1:J2:near-horizon:M3}
 ds^2_{\cM_3} & = ds^2_{S^3}
 - \left[ \Gamma^2 \gamma_3^{1/3} - \frac{a g (4 + 5 a g)}{\Xi} \right] d\gamma^2
 + 2 R_{\text{AdS}_2} \, \Gamma \tilde r \, d\tilde t \, d\gamma \, , \\
 \Gamma & = \frac{g \left( 3 a^4 + 4 a^2 r_m^2 + \beta_2 \right)}{2 \Xi^2 \gamma_3^{2/3}} \, ,
\eea
and
\bea
 ds^2_{S^3} =  \frac14 \left( d\vartheta^2 + d\varphi^2 + d\chi^2 + 2 \cos \vartheta \, d\varphi \, d\chi \right) = \frac14 \sum_{i = 1}^{3} \sigma_i \, , \qquad \qquad
 d\gamma = \frac{\sigma_3}{2} \, .
\eea
Here, $\sigma_i$ $(i = 1 , 2, 3)$ are the right-invariant $\SU(2)$ one-forms,
\bea
 \sigma_1 & = - \sin \chi \, d\vartheta + \cos \chi \sin \vartheta \, d\varphi \, , \\
 \sigma_2 & = \cos \chi \, d\vartheta + \sin \chi \sin \vartheta \, d\varphi \, , \\
 \sigma_3 & = d\chi + \cos \vartheta \, d\varphi \, .
\eea
Notice that
\be
 ds^2_{S^2} = \sigma_1^2 + \sigma_2^2 = d\vartheta^2 + \sin^2 \vartheta \, d \varphi^2 \, .
\ee
Due to the constraint \eqref{gravity:constraint} we can simplify
\bea
 \frac{a (4 + 5 a g)}{\Xi} = g \gamma_1 \, , \qquad \qquad
 \Gamma = \frac{g \gamma_2}{2 \gamma_3^{2/3}} \, .
\eea
Upon a further rescaling of the time coordinate
\bea
 \tilde t = - \frac{1}{2} \sqrt{4 - \frac{g^2 \gamma_2^2}{\gamma_3 (1 + g^2 \gamma_1)}} \, \tau \, ,
\eea
the near-horizon metric with squashed AdS$_2 \times_w S^3$ geometry and the gauge fields can be brought to the form:
\bea
 \label{J1:J2:final:near-horizon}
 ds^2_{(5)} & = R^2_{\text{AdS}_2}\ ds^2_{\text{AdS}_2}
 + \frac{R_{S^2}^2}{4} \left[ ds^2_{S^2} + \upsilon \left( \sigma_3 - \alpha \, \tilde r d\tau \right)^2 \right] \, , \\
 L^I & = \frac{\gamma_3^{1/3}}{\mu_I} \, , \qquad \qquad
 A^I_{(5)} = e^I \, \tilde r \, d \tau - f^I \, \sigma_3 \, .
\eea
Here, we defined the constants
\bea
 \alpha & = \frac{g \gamma_2}{2 \left( 1 + g^2 \gamma_1 \right)  \sqrt{\gamma_3 \upsilon}} \, , \qquad \qquad
 && R_{S^2}^2 = \gamma_3^{1/3} \, , \\
 e^I & = - \frac{\sqrt{\gamma_3 \upsilon}}{2 \mu_I (1 + g^2 \gamma_1) } \, ,
 && f^I = \frac{g}{4} ( \mu_I - \gamma_1 ) + \frac{g \gamma_3}{4 \mu_I^2} \, ,
\eea
and
\bea
 \upsilon = 1 + g^2 \gamma_1 - \frac{g^2 \gamma_2^2}{4 \gamma_3} \, .
\eea
Note that we added the subscript $(5)$ in order to emphasize that these are five-dimensional quantities
which will next be related to a solution in four dimensions via dimensional reduction along the $\chi$ direction.

\subsection{Dimensional reduction on the Hopf fibres of squashed $S^3$}
\label{ssec:reduction along S3}

In five-dimensional supergravity theories, including $n_{\rm V}$ abelian gauge fields $A_{(5)}^I$
and real scalar fields $L^I$ $(I = 1 , \ldots , n_{\rm V})$ coupled to gravity, the rules for reducing the bosonic fields are the following \cite{Andrianopoli:2004im,Behrndt:2005he,Cardoso:2007rg,Aharony:2008rx,Looyestijn:2010pb}:
\bea
 \label{reduction:rules:5dBH}
 ds_{(5)}^2 & = e^{2 \phi} \, ds_{(4)}^2 + e^{- 4 \phi} \, \left( dx^5 - A_{(4)}^0 \right)^2 \, , \qquad && dx^5 = d\chi \, , \\
 A_{(5)}^I & = A_{(4)}^I + \re z^I \left( dx^5 - A_{(4)}^0 \right) \, , \\
 L^I & = e^{2 \phi} \im z^I \, , && e^{-6 \phi} = \frac16 C_{I J K}  \im z^I  \im z^J  \im z^K \, ,
\eea
where $ds_{(4)}^2$ denotes the four-dimensional line element,
the $A_{(4)}^\Lambda$ $( \Lambda = 0, I )$ are the four-dimensional abelian gauge fields and $z^I$
are the complex scalar fields in four dimensions. Our conventions  for $\cN=2$ gauged supergravity in four dimensions
are presented in appendix \ref{app:4d gauged sugra}. 
The four-dimensional theory has $n_{\rm V}$ abelian vector multiplets,
parameterizing a special K\"ahler manifold $\cM$ with metric $g_{i \bar{j}}$,
in addition to the gravity multiplet (thus a total of $n_{\rm V}+1$ gauge fields and $n_{\rm V}$ complex scalars).
The scalar manifold is defined by the prepotential $\cF \left( X^\Lambda \right)$,
which is a homogeneous holomorphic function of sections $X^\Lambda$,
\bea
 \cF \left( X^\Lambda \right) = - \frac16 \frac{C_{I J K} X^I X^J X^K}{X^0} = - \frac{X^1 X^2 X^3}{X^0} \, .
\eea 
In the second equality we employed the five-dimensional supergravity data for the STU model from section \ref{sec:AdS5 black holes}.
In $\cN = 2$ gauged supergravity in four dimensions the $\U(1)_R$ symmetry, rotating the gravitini,
is gauged by a linear combination of the (now four) abelian gauge fields.
The coefficients are called Fayet-Iliopoulos (FI) parameters $g_\Lambda$
and three of them can be directly read off the five-dimensional theory: $g_1 = g_2 = g_3 = 1$.\footnote{In consistent models one can always apply an
electric-magnetic duality transformation so that the corresponding gauging becomes purely electric, \ie, $g^\Lambda = 0$.}
The last coefficient, $g_0$, measuring how the Kaluza-Klein gauge potential $A^0_{(4)}$
enters the R-symmetry, can be left arbitrary for the moment.
This can be achieved by a Scherk-Schwarz reduction when allowing a particular reduction ansatz for the gravitino
as explained in \cite{Andrianopoli:2004im,Andrianopoli:2004xu,Andrianopoli:2005jv,Looyestijn:2010pb,Hristov:2014eba}.
The prepotential and the FI parameters uniquely specify the four-dimensional $\cN = 2$ gauged supergravity Lagrangian and BPS variations.

Now, we can proceed with the explicit reduction of the line element \eqref{J1:J2:final:near-horizon} on the Hopf fibres of $S^3$ viewed as a $\U(1)$ bundle over $S^2 \cong \bC\bP^1$.
We thus identify $x^5$ with $\chi$. The four-dimensional solution takes the form\footnote{We have rescaled the time coordinate,
$\tilde{\tau} \equiv - R_{S^2} R_{\text{AdS}_2} \sqrt{\upsilon} \, \tau / 2$,
in order to put the AdS$_2$ part of the metric in the canonical coordinates.}
\bea
\label{eq:4dsolution}
 ds_{(4)}^2 & = - e^{2 U} \, d\tilde{\tau}^2 + e^{-2 U} dr^2+ e^{2 \left( V - U \right)}
 \left( d\vartheta^2 + \sin^2\vartheta \, d\varphi^2 \right) \, , \\
 A_{(4)}^0 & = \tilde{q}_{(4)}^0(r) \, d\tilde{\tau} - \cos\vartheta \, d\varphi \, ,
 \qquad \qquad A_{(4)}^I = \tilde{q}_{(4)}^I(r) \, d\tilde{\tau} \, ,
\eea
where
\bea
 e^{U} & = \frac{\sqrt{2}}{R_{\text{AdS}_2} R_{S^2}^{1/2} \upsilon^{1/4}} \, r \, ,
 && e^{V} = \frac{R_{S^2}}{2 R_{\text{AdS}_2}} \, r \, , \\
 \tilde{q}_{(4)}^0(r) & = - \frac{2 \alpha}{R^2_{\text{AdS}_2} R_{S^2} \upsilon^{1/2}} \, r \, ,
 \qquad && \tilde{q}_{(4)}^I(r) = - \frac{2 \left( e^I - f^I \alpha \right)}{R^2_{\text{AdS}_2} R_{S^2} \upsilon^{1/2}} \, r \, .
\eea
The complex scalars are given by
\bea
 \label{4d:physical:scalars}
 z^I = - f^I + \frac{i}{2} R_{S^2} \upsilon^{1/2} L^I
 = - f^I + \frac{i}{2} \frac{\upsilon^{1/2} \gamma_3^{1/2}}{\mu_I} \, .
\eea
Employing Eq.\,\eqref{em:charges:4d:scalars} we can compute the conserved electric charges.
After some work they read
\bea
 \label{J1:J2:em:charges:4d:final}
 q_{0} & = \frac{g}{8} \left( \gamma_2 + 2 g^2 \gamma_3 \right)
 = \frac{G_{\text{N}}^{(5)}}{\pi} J_{\phi} \, , \\
 q_I & = - \frac{1}{4} \left[ \mu_I + \frac{g^2}{2} \left( \gamma_2 - \frac{2 \gamma_3}{\mu_I} \right) \right] 
 = - \frac{G_{\text{N}}^{(5)}}{\pi} Q_I \, .
 \eea
This is in agreement with \cite{Astefanesei:2011pz}.
The magnetic charges of the four-dimensional solution can be directly read off the spherical components of the gauge fields $A^\Lambda_{(4)}$ \eqref{eq:4dsolution},
\bea
 p^0 = 1 \, , \qquad \qquad p^I = 0 \, .
\eea
The entropy of the four-dimensional black hole precisely equals the entropy of the rotating black hole in five dimensions,
\bea
 \label{4d:entropy:5d}
 S_{\text{BH}}^{(4)} = \frac{\text{Area}^{(4)}}{4 G^{(4)}_{\rm N}} = \frac{\pi e^{2 \left( V - U \right)}}{G^{(4)}_{\rm N}} = \frac{\pi^2 R_{S^2}^3 \upsilon^{1/2}}{2 G^{(5)}_{\rm N}} = S_{\text{BH}}^{(5)} \, ,
\eea
upon using the standard relation 
\bea
 \frac{1}{G_{\rm N}^{(4)}} = \frac{4 \pi}{G_{\rm N}^{(5)}}\, .
\eea

\subsection{Attractor mechanism in four dimensions}
\label{ssec:the attractor mechanism}

Let us define the central charge of the black hole $\cZ$ and the superpotential $\cL$,
\bea
 \cZ = e^{\cK / 2} \left( q_\Lambda X^\Lambda - p^\Lambda \cF_\Lambda \right) \, , \qquad \qquad
 \cL = e^{\cK / 2} \left( g_\Lambda X^\Lambda - g^\Lambda \cF_\Lambda \right) \, .
\eea
The BPS equations for the near-horizon solution \eqref{ansatz:metric:gauge field:4d}
with constant scalar fields $z^i$ imply that \cite{DallAgata:2010ejj}:\footnote{From comparing \eqref{attractor:initial:4d} with equations (3.5) and (3.8) in \cite{DallAgata:2010ejj},
we see that they differ by a factor of $2$. This is due to our different convention for the action (see footnote \ref{convention:action:metric}).}
\bea
 \label{attractor:initial:4d}
 \cZ + 2 i e^{2 \left( V - U \right)} \cL = 0 \, , \qquad \qquad
 D_j \left( \cZ + 2 i e^{2 \left( V - U \right)} \cL \right) = 0 \, ,
\eea
which can be rewritten as
\bea
 \label{attractor:4d}
 \partial_j \frac{\cZ}{\cL} = 0 \, , \qquad \qquad
 i \frac{\cZ}{\cL} = 2 e^{2 \left( V - U \right)} \, .
\eea
Here, $D_j = \partial_j + \frac12 \partial_j \cK$ with $\cK$ being the K\"ahler potential [see Eq.\,\eqref{4d:Kahler:prepotential}].
Therefore, the complex scalar fields $z^i$ are fixed at the horizon such that the quantity
$i \frac{\cZ}{\cL}$ has a critical point on $\cM$ and then its value is proportional to the Bekenstein-Hawking entropy of the BPS black hole.

We can extremize the quantity $i \frac{\cZ}{\cL}$ under the following gauge fixing constraint, which precisely corresponds to \eqref{constraint:N=4},
\bea
 \label{eq:gaugefixing}
 g_0 X^0 + \sum_{I = 1}^{3} X^I = 1 \, ,
\eea
where we plugged in the explicit values for the FI parameters, \ie, $g_1 = g_2 = g_3 = 1$.
The real sections $X^\Lambda$ are constrained in the range $0 < X^\Lambda < 1$.
We find that
\bea
 \label{J1:J2:attractor:equations}
 \partial_I \left[ \sum_{I = 1}^{3} X^I \left( q_I - \frac{q_0}{g_0} \right) + \frac{q_0}{g_0}
 - \frac{g_0^2 X^1 X^2 X^3}{\left( 1 - X^1 - X^2 - X^3 \right)^2} \right] = 0 \, , \quad \text{ for } \quad I = 1 , 2 , 3 \, ,
\eea
where $\partial_I \equiv \partial / \partial X^I$.
The sections at the horizon are obtained from
\bea
 X^{0} = \frac{1}{g_0 \left( 1 + z^1 + z^2 + z^3 \right)} \, , \qquad \qquad X^{1,2,3} = \frac{z^{1,2,3}}{1 + z^1 + z^2 + z^3} \, .
\eea

We are now in a position to determine the value of the  FI parameter $g_0$.
Partial topological A-twist along $S^2$, embedding the spacetime holonomy group into the R-symmetry,
ensures that $\cN = 2$ supersymmetry is preserved in four dimensions \cite{Witten:1991zz}.
The twisting amounts to an identification of the spin connection with the R-symmetry
so that one of the SUSY parameters becomes a scalar.
This leads to the following Dirac-like quantization condition \cite{Cacciatori:2009iz,DallAgata:2010ejj,Hristov:2010ri}:
\bea
 g_{\Lambda} p^{\Lambda}  = 1 = g_0 p^0 \, ,
\eea
which fixes the value of $g_0 = 1$.
It is straightforward to check that, substituting the values for the physical scalars at the horizon \eqref{4d:physical:scalars},
the charges \eqref{J1:J2:em:charges:4d:final}, and setting $g_0 = 1$, Eq.\,\eqref{J1:J2:attractor:equations} is fulfilled.
The scalars $\bar{z}^i(r_{\rm h})$ at the horizon are determined in terms of the black hole charges $q_I$ by virtue of the attractor equations:
\bea
 q_I - q_0 = \left( 2 + \frac{1}{\bar{z}^I} \right) \bar{z}^1 \bar{z}^2 \bar{z}^3 \, .
\eea
The value of $i \frac{\cZ}{\cL}$ at the critical point yields,
\bea
 i \frac{\cZ}{\cL} \bigg|_{\rm crit} \left( q_\Lambda \right) = 2 e^{2 \left( V - U \right)}
 = \frac{2 G^{(4)}_{\rm N}}{\pi} S^{(4)}_{\rm BH} \left( q_\Lambda \right)
 \, .
\eea
The holding of the four-dimensional BPS attractor mechanism for
the dimensionally reduced near-horizon geometry \eqref{eq:4dsolution}
proves that the dimensional reduction preserves the full
amount of supersymmetries originally present in five dimensions.

Due to the very suggestive form of the attractor equations \eqref{J1:J2:attractor:equations}
it is now not hard to compare them with the five-dimensional extremization.

\subsection{Comparison with five-dimensional extremization}
\label{ssec:ESUSY:extremization}

Consider the quantity $E$ in \eqref{eq:Casimir:rewrittenX}
rewritten in terms of the chemical potentials for $J^{\pm}$ and $Q_{1,2,3}$.
Recall that we are focusing on the case with equal angular momenta, \ie, $J_{\psi} = J_{\phi}$ (so $J^{-} = 0)$.
Extremizing \eqref{rotating:attractor:4d} with respect to $X^0_{-}$ fixes the value of $X^0_{-} = 0$.
Thus, the black hole entropy is obtained by extremizing the quantity
\bea
 \label{static:attractor:4d}
 \cS \big|_{J^- = 0} =   \frac{2 \pi^2 i}{G_{{\rm N}}^{(5)}} \frac{X^1 X^2 X^3}{\left( X^0_{+} \right)^2} + 2 \pi i \sum_{I = 1}^{3} Q_I X^I  - \pi i J^{+} X_{+}^0 
\, ,
\eea
subject to the constraint \eqref{eq:constraintX}.
Identifying $X^{0}$ in \eqref{J1:J2:attractor:equations}
with $X^{0}_{+}$ in \eqref{static:attractor:4d},
and using $g_0 = 1$ together with \eqref{J1:J2:em:charges:4d:final},
we find that the extremization of $\cS$ 
corresponds to the four-dimensional attractor mechanism on the gravity side and they lead
to the same entropy.

\section{Discussion and future directions}
\label{sec:discussion}

We have shown that the entropy of a supersymmetric rotating black hole in AdS$_5$ with electric charges $Q_I$ $(I = 1, 2 , 3)$
and angular momenta $J_\phi \equiv J_1$, $J_\psi \equiv J_2$ can be obtained as the Legendre
transform of the quantity $(-E)$ in \eqref{modified:Casimir:SUSY:N=4},
\be\label{E-extremization}
 \cS(Q_I,J_i) =  - E(\Delta_I,\omega_i)  +  2 \pi i  \bigg( \sum_{I=1}^3 Q_I \Delta_I -  \sum_{i = 1}^2 J_i \omega_i \bigg) \bigg |_{\bar\Delta_I, \bar\omega_i} \, ,
\ee
where $\bar\Delta_I$ and  $\bar\omega_i$ are the extrema of the functional on the right hand side.

The result is quite intriguing and deserves a better explanation and understanding. We leave a more careful analysis for the future. For the moment, let us just make few preliminary observations.

The quantity $E$ can be interpreted  as a combination of 't Hooft anomaly polynomials that arise studying  the partition function $Z_{{\cal N}=4}(\Delta_I,\omega_i)$ on $S^3\times S^1$
or the superconformal index $I(\Delta_I,\omega_i)$ for ${\cal N}=4$ SYM \cite{Bobev:2015kza,Brunner:2016nyk}.
Some explicit expressions are given in appendix \ref{AppC}.
In particular, $E$ is  {\it formally} equal to the supersymmetric Casimir energy of ${\cal N}=4$ SYM as a function of the chemical potentials (see for example equation (4.50) in \cite{Bobev:2015kza} and appendix \ref{AppC}).
However, this analogy is only formal since we are imposing the constraint \eqref{constraint:N=4}. Chemical potentials are only defined modulo 1, so the constraint to be imposed on them also suffers from angular ambiguities.
Consistency of the index and partition function just requires  $\sum_{I=1}^3 \Delta_I +\sum_{i=1}^2 \omega_i \in \mathbb{Z}$. To recover the known expressions for the supersymmetric Casimir energy and for consistency with gauge anomaly cancellations \cite{Assel:2015nca,Bobev:2015kza},
one needs to impose  $\sum_{I=1}^3 \Delta_I +\sum_{i=1}^2 \omega_i =0$, and this contrasts with \eqref{constraint:N=4}. 

It would be tempting to interpret  the Legendre transform \eqref{E-extremization} as a result of the saddle-point approximation
of a Laplace integral of $Z_{{\cal N}=4}$ in the limit of large charges (large $N$).\footnote{We are ignoring here potential sign ambiguities in the definition of charges.}
Ignoring angular ambiguities, $E$ is  the leading contribution at order $N^2$
of the logarithm of the partition function $Z_{{\cal N}=4}$ on $S^3\times S^1$.
Indeed, $\log Z_{{\cal N}=4}= - E + \log I$ \cite{Assel:2014paa,Lorenzen:2014pna,Assel:2015nca,Bobev:2015kza,Martelli:2015kuk,Genolini:2016sxe,Brunner:2016nyk}
and the index is a quantity independent of $N$ for generic values for the chemical potentials  \cite{Kinney:2005ej}.
In these terms, the result would be completely analogous to the connection between asymptotically
AdS$_4\times S^7$ black hole entropy and the topologically twisted index of ABJM \cite{Benini:2015eyy,Benini:2016rke}.

The appearance of the supersymmetric Casimir energy can be surprising since the entropy counts the degeneracy of ground states of the system. 
However, the dimensional reduction to four dimensions performed in  Section \ref{sec:a limiting case} offers a different perspective on this point. The dimensionally reduced black hole
is static but not asymptotically AdS. Let us assume that we can still use holography. In the dimensional reduction,  a magnetic flux $p^0$ is turned on for the graviphoton. 
This means that supersymmetry is preserved with a topological twist.  The same should be true for the boundary theory.  It is then tempting to speculate that, upon dimensional reduction,  the partition function $Z_{{\cal N}=4}$ becomes the topologically twisted index of the boundary three-dimensional theory \cite{Benini:2015noa}. The supersymmetric Casimir energy, which is the leading contribution  of $\log Z_{{\cal N}=4}$ at large $N$ then becomes the leading contribution of the three-dimensional topologically twisted index and the latter is known to correctly account  for the microstates of four-dimensional black holes \cite{Benini:2015eyy,Benini:2016rke}.

The above discussion ignores completely the angular ambiguities and the role of the  constraint \eqref{constraint:N=4},
which should be further investigated.   For sure, the result of the extremization of $E$ is complex and lies in the region
where the chemical potentials satisfy \eqref{constraint:N=4}. Unfortunately, we are not aware of a general discussion
of the possible regularizations  of $Z_{{\cal N}=4}$ that takes into account the angular ambiguities. Moreover, there is some recent claim \cite{Papadimitriou:2017kzw,An:2017ihs} 
of the presence of an anomaly in the supersymmetry transformations leading to a modification of the BPS condition in gravity
that would be interesting to investigate further in this context. 

Both the constraint \eqref{constraint:N=4} and the analogous of the more traditional one $\sum_{I=1}^3 \Delta_I +\sum_{i=1}^2 \omega_i =0$
have been used in the literature to explore different features of $Z_{{\cal N}=4}$ or the index.
The traditional constraint has been used in the analysis of the high temperature limit of the index
\cite{Ardehali:2015hya,Ardehali:2015bla} (see also \cite{Shaghoulian:2016gol,DiPietro:2016ond}) and in the study of factorization properties \cite{Nieri:2015yia}.
On the other hand, the importance of \eqref{constraint:N=4} has been stressed in  \cite{Brunner:2016nyk}
where the constraint has been used to extract the supersymmetric Casimir energy directly from the superconformal index.\footnote{Interestingly, the same constraint is also used in relating
the universal part of supersymmetric R\'enyi entropy to an equivariant integral of the anomaly polynomial \cite{Yankielowicz:2017xkf}.} See appendix \ref{AppC} for more details.
In the low temperature limit, which can be obtained by rescaling $\Delta_I \to \beta\Delta_I, \omega_i \to \beta \omega_i$ and taking large $\beta$, the angular ambiguity in the constraint disappears.
 
Finally, it is worth mentioning that angular ambiguities also played a prominent role  in the evaluation of the saddle-point for the topologically twisted index of the ABJM theory and the comparison with the entropy of AdS$_4$ black holes \cite{Benini:2015eyy}.

All this is quite speculative and we hope to come back with more precise statements in the future. 
There are also other directions to investigate. Let us mention some of them.

\paragraph*{Attractor mechanism in five dimensions ---}
it is known that in five-dimensional $\cN = 2$ \emph{ungauged} supergravity
the near-horizon geometry of an extremal BPS black hole is
governed by the attractor mechanism \cite{Ferrara:1996dd,Ferrara:1996um,Chamseddine:1996pi,Kallosh:1996vy}.
That is, the values of the scalar fields at the horizon are fixed by black hole charges, and
the area of the black hole horizon is given in terms of the extremal value of the central charge $\bar\cZ$ in moduli space
and the angular momentum $J$,
\bea
 \text{Area}^{(5)} =\frac{\pi^2}{3}\sqrt{\bar\cZ^3 - J^2} \, .
\eea
It would be interesting to find analogous results in five-dimensional \emph{gauged} supergravity\footnote{First-order flow equations for stationary black brane solutions
and magnetically charged black strings in five-dimensional $\cN =2$ gauged supergravity were analyzed in \cite{BarischDick:2012gj,Klemm:2016kxw,Amariti:2016mnz}.}
and see if we can recover the extremization principle \eqref{E-extremization}.

\paragraph*{Rotating attractors in four dimensions ---}
as explained in section \ref{sec:a limiting case}
we only understand well the static attractor equations in four dimensions.
It is natural to extend this analysis to rotating cases, which will correspond
to the refinement by angular momentum in the dual field theory.
We would have already had an example of a rotating attractor if we were to consider
the dimensional reduction of the generic BPS black hole with $J_{\psi} \neq J_{\phi}$.

\paragraph*{Dimensional reduction of black strings in five dimensions ---}
similarly to the reduction of BPS black holes from five to four dimensions,
one could perform a reduction between five-dimensional BPS black strings
and four-dimensional BPS black holes, as it was already done in \cite{Hristov:2014eza}.
Given the recent results \cite{Hosseini:2016cyf} on the topologically twisted index
for $\cN = 1$ supersymmetric theories on $S^2 \times T^2$,
this could imply a relation between the $c$-extremization of the
two-dimensional SCFTs \cite{Benini:2012cz,Benini:2013cda}
and the $\cI$-extremization principle of \cite{Benini:2015eyy,Benini:2016rke}.

\section*{Acknowledgements}
We would like to thank Arash Arabi Ardehali, Francesco Benini, Nikolay Bobev, Sara Pasquetti,  Vyacheslav S. Spiridonov and Chiara Toldo for useful discussions and comments.
AZ is partially supported by the INFN and ERC-STG grant 637844-HBQFTNCER.
SMH is supported in part by INFN. KH is supported in part by the Bulgarian NSF grant DN08/3.

\appendix

\section{Five-dimensional $\cN = 2$ gauged supergravity}
\label{app:5d gauged sugra}

The theory we shall be considering, following the conventions of \cite{Looyestijn:2010pb}, is the five-dimensional $\cN = 2$ Fayet-Iliopoulos (FI) gauged supergravity
coupled to $n_{\rm V}$ vector multiplets. It is based on a homogeneous real cubic polynomial
\bea
 \cV \left( L^I \right) = \frac16 C_{I J K} L^I L^J L^K \, ,
\eea
where $I, J , K = 1, \ldots , n_{{\rm V}}$ and $C_{I J K}$ is a fully symmetric third-rank tensor appearing in the Chern-Simons term.
Here, $L^I(\varphi^i)$ are real scalars satisfying the constraint $\cV=1$.
The action for the bosonic sector reads
\bea
 S^{(5)} = \int_{\bR^{4,1}} & \bigg[ \frac12 R^{(5)} \star_5 1- \frac12 G_{I J} \rd L^I \wedge \star_5 \rd L^J - \frac12 G_{I J} F^I \wedge \star_5 F^J \\
 & - \frac{1}{12} C_{I J K} F^I \wedge F^J \wedge A^K + \chi^2 V \star_5 1 \bigg] \, ,
\eea
where $R^{(5)}$ is the Ricci scalar, $F^I \equiv \rd A^I$ is the Maxwell field strength, and $G_{I J}$ can be written in terms of $\cV$,
\bea
 \label{metric:gauge:kinetic:V}
 G_{I J} = - \frac12 \partial_I \partial_J \log \cV \big|_{\cV = 1} \, .
\eea
We also set $8 \pi G^{(5)}_{\text{N}} = 1$.
Furthermore, it is convenient to define
\bea
 \label{real:sections}
 L_I \equiv \frac16 C_{I J K} L^J L^K \, .
\eea
Therefore, we find that
\bea
 \label{metric:gauge:kinetic:sections}
 G_{I J} = \frac92 L_I L_J- \frac12 C_{I J K} L^K \, , \qquad \qquad L^I L_I = 1 \, ,
\eea
and
\bea
 L_I = \frac{2}{3} G_{I J} L^J \, , \qquad \qquad L^I = \frac{3}{2} G^{I J} L_J \, ,
\eea
where $G_{I K} G^{K J} = \delta^I_J$.
The inverse of $G_{I J}$ is given by
\bea
 G^{I J} = 2 L^I L^J - 6 C^{I J K} L_K \, ,
\eea
where $C^{I J K} \equiv C_{I J K}$. We then have
\bea
 L^I = \frac92 C^{I J K} L_J L_K \, .
\eea
The metric on the scalar manifold is defined by
\bea
 g_{i j} = \partial_i L^I \partial_j L^J G_{I J} \big|_{\cV = 1} \, ,
\eea
and the scalar potential reads
\bea
 V (L) = V_I V_J \left(6 L^I L^J - \frac92 g^{i j} \partial_i L^I \partial_j L^J \right) \, .
\eea
Here, $V_I$ are FI constants which are related to the vacuum value $\bar L_I$ of the scalars $L_I$,
\be
 \bar L_I = \xi^{-1} V_I \, ,
\ee
where $\xi^3 = \frac92 C^{I J K} V_I V_J V_K$ and the AdS$_5$ radius of curvature is given by $g^{-1} \equiv \left( \xi \chi \right)^{-1}$.
A useful relation of very special geometry is,
\bea
 g^{i j} \partial_i L^I \partial_j L^J = G^{I J} - \frac23 L^I L^J \, .
\eea
Thus,
\bea
 V(L) = 9 V_I V_J \left( L^I L^J - \frac12 G^{I J} \right) \, .
\eea

\section{Four-dimensional $\cN = 2$ gauged supergravity}
\label{app:4d gauged sugra}

We consider the four-dimensional $\cN = 2$ FI-gauged supergravity with a holomorphic prepotential
\bea
 \label{4d:prepotential:cubic}
 \cF \left( X^\Lambda \right) = - \frac16 \frac{C_{I J K} X^I X^J X^K}{X^0} \, ,
\eea
where $X^\Lambda$ represent the symplectic sections and $\Lambda = 0, 1 , 2 , 3$.
The physical scalars $z^I$ $(I = 1 , 2 , 3)$ belonging to the vector multiplets are given by
\bea
 z^I = \frac{X^I}{X^0} \, ,
\eea
and parameterize a special K\"ahler manifold $\cM$ of complex dimension $n_{\rm V}$ with metric
\bea
 \label{Kahler:metric:Kahler}
 g_{i \bar{j}} = \partial_i \partial_{\bar{j}} \cK(z , \bar{z}) \, .
\eea
Here, $\cK(z , \bar{z})$ is the K\"ahler potential and it reads
\bea
 \label{4d:Kahler:prepotential}
 e^{- \cK (z , \bar{z})} = i \left( \bar{X}^\Lambda \cF_\Lambda - X^\Lambda \bar{\cF}_\Lambda \right) \, ,
\eea
where $\cF_\Lambda \equiv \partial_\Lambda \cF$.
Plugging \eqref{4d:prepotential:cubic} into \eqref{4d:Kahler:prepotential} we find that
\bea
 \label{4d:Kahler:scalars}
 e^{- \cK (z , \bar{z})}  = \frac{4 i}{3} C_{I J K} \im z^I \im z^J \im z^K = 8 e^{- 6 \phi} \, ,
\eea
where due to the symmetries of the theory we can set $X^0 = 1$.
In the last equality we used Eq.\,\eqref{reduction:rules:5dBH}.
Substituting \eqref{4d:Kahler:scalars} into \eqref{Kahler:metric:Kahler} and using
\bea
 \frac{\partial}{\partial z^I} & = \frac12 \left( \frac{\partial }{\partial \re z^I} - i \frac{\partial}{\partial \im z^I} \right) \, , \qquad \qquad
 \frac{\partial}{\partial \bar{z}^I} & = \frac12 \left( \frac{\partial }{\partial \re z^I} + i \frac{\partial}{\partial \im z^I} \right) \, ,
\eea
we can write the K\"ahler metric as
\bea
 \label{Kahler:metric:scalars}
 g_{I J} & = - \frac14 \frac{\partial}{\partial \im z^I} \frac{\partial}{\partial \im z^J}
 \log \left( \frac{4i}{3} C_{I J K} \im z^I \im z^J \im z^K \right) \\
 & = - \frac{1}{4 e^{- 6 \phi}} \left( C_{I J} - \frac{C_I C_J}{4 e^{- 6 \phi}} \right) \, .
\eea
Here, we introduced the following notation
\bea
 C_{I J} = C_{I J K} \im z^K \, , \qquad \qquad C_I = C_{I J K} \im z^J \im z^K \, .
\eea
The action of the bosonic part of the supergravity reads \cite{Andrianopoli:1996cm,Louis:2002ny}\footnote{\label{convention:action:metric}We follow
the conventions of \cite{Looyestijn:2010pb}, which is different from \cite{Andrianopoli:1996cm}
by factors of two in the gauge kinetic terms and the scalar potential $V(z, \bar{z} )$.
One can swap between the conventions by rescaling the four-dimensional metric $g_{\mu \nu} \to \frac12 g_{\mu \nu}$
and then multiplying the action by $2$. This will modify the definition of the symplectic-dual gauge field strength $G_\Lambda$ by a factor of 2, see Eq.\,\eqref{symplectic-dual:F}.}
\bea
 S^{(4)} = \int_{\bR^{3,1}} & \bigg[ \frac{1}{2} R^{(4)} \star_4 1 + \frac14 \im \cN_{\Lambda \Sigma} F^\Lambda \wedge \star_4 F^{\Sigma}
 + \frac14 \re \cN_{\Lambda \Sigma} F^{\Lambda} \wedge F^{\Sigma} \\
 & - g_{i \bar{j}} D z^{i} \wedge \star_4 D \bar{z}^{\bar{j}}
 - V(z, \bar{z}) \star_4 1 \bigg] \, ,
\eea
where $\Lambda, \Sigma = 0 , 1 , \ldots , n_{\rm V}$ and $i , \bar{j}= 1 , \ldots , n_{\rm V}$.
Note that we already set $8 \pi G_{{\rm N}}^{(4)} = 1$.
Here, $V$ is the scalar potential of the theory,
\bea
 V(z, \bar{z}) = 
 2 g^2 \left( U^{\Lambda \Sigma} - 3 e^{\cK} \bar{X}^\Lambda X^\Sigma \right) \xi_\Lambda \xi_\Sigma \, ,
\eea
where $\xi_\Lambda$ are the constant quaternionic moment maps known as FI parameters and
\bea
 U^{\Lambda \Sigma} = - \frac{1}{2} \left( \im \cN \right)^{-1 | \Lambda \Sigma } - e^{\cK} \bar{X}^\Lambda X^\Sigma \, .
\eea
The matrix $\cN_{\Lambda \Sigma}$ of the gauge kinetic term is a
function of the vector multiplet scalars and is given by
\bea
\label{period:matrix:4d}
 \cN_{\Lambda \Sigma} = \bar{\cF}_{\Lambda \Sigma} + 2 i \frac{\im \cF_{\Lambda \Delta} \im \cF_{\Sigma \Theta} X^\Delta X^\Theta}{\im \cF_{\Delta \Theta} X^\Delta X^\Theta} \, .
\eea

In this paper we focus on black holes with the near-horizon geometry AdS$_2 \times S^2$.
The ansatz for the metric and gauge fields is
\bea
 \label{ansatz:metric:gauge field:4d}
 ds^2 & = - e^{2 U} \, d\tau^2 + e^{-2 U} \, dr^2 + e^{2 \left( V - U \right)} \left( d\vartheta^2 + \sin^2\vartheta \, d\varphi^2 \right) \, , \\
 A^\Lambda & = \tilde{q}^\Lambda(r) \, d\tau - p^\Lambda(r) \cos\vartheta \, d\varphi \, .
\eea
The black hole magnetic and electric charges are then given by \cite{DallAgata:2010ejj,Halmagyi:2013sla}
\bea
 \label{em:charges:4d:period}
 p^\Lambda & \equiv \frac{1}{4 \pi} \int_{S^2} F^\Lambda \, , \\
 q_\Lambda & \equiv \frac{1}{4 \pi} \int_{S^2} G_\Lambda
 = e^{2 \left( V - U \right)} \im \cN_{\Lambda \Sigma} \, \tilde{q}'^{\Sigma}
 + \re \cN_{\Lambda \Sigma} \,  p^{\Sigma} \, ,
\eea
where we defined the symplectic-dual gauge field strength,
\be
 \label{symplectic-dual:F}
 G_\Lambda \equiv 2 \frac{\delta \cL}{\delta F^{\Lambda}}
 =  \im \cN_{\Lambda \Sigma} \star_4 F^{\Sigma}
 + \re \cN_{\Lambda \Sigma} F^{\Sigma} \, ,
\ee
such that $\left( F^\Lambda , G_\Lambda \right)$ transforms as a $\left( 2 , n_{\rm V} + 2 \right)$
symplectic vector under electric-magnetic duality transformations.
Employing Eq.\,\eqref{period:matrix:4d} we find that
\bea
 \label{period:matrix:4d:scalars}
 \im \cN_{I J} & = \cG_{I J} \, , && \re \cN_{I J} = - C_{I J K} \re z^K \, ,\\
 \im \cN_{I 0} & = - \cG_{I J} \re z^J \, , && \re \cN_{I 0} = + \frac12 C_{I J K} \re z^J \re z^K \, ,\\
 \im \cN_{0 0} & = - \left( e^{-6 \phi} - \cG_{I J} \re z^I \re z^J \right) \, , \qquad && \re \cN_{0 0} = - \frac13 C_{I J K} \re z^I \re z^J \re z^K  \, ,
\eea
where we defined
\bea
 \cG_{I J} \equiv C_{I J} - \frac{C_I C_J}{4 e^{- 6 \phi}} \, .
\eea
Therefore, the electric charges read
\bea
 q_{0} & = - e^{2 \left( V - U \right)} \tilde{q}'^{0}
 \left[ e^{- 6 \phi} + \cG_{I J} \re z^J \left( \frac{\tilde{q}'^{I}}{\tilde{q}'^{0}} - \re z^I \right) \right] \\
 & + \frac{p^{0}}{2} C_{I J K} \re z^J \re z^K \left( \frac{p^I}{p^0} - \frac23 \re z^I \right) \, , \\
 q_{I} & = e^{2 \left( V - U \right)} \tilde{q}'^{0} \cG_{I J} \left( \frac{\tilde{q}'^{J}}{\tilde{q}'^{0}} - \re z^J \right)
 - p^{0} C_{I J K} \re z^K \left( \frac{p^J}{p^0} - \frac12 \re z^J \right) \, .
\eea

In the gauged STU model $(n_{\rm V} = 3)$ the only nonvanishing intersection numbers are $C_{1 2 3} = 1$ (and cyclic permutations).
Hence,
\bea
 \cG_{I J} =
 \begin{cases}
 - \frac{\im z^1 \im z^2 \im z^3}{ \left( \im z^I \right)^2} &\text{if } I = J \\
 0 &\text{otherwise}
 \end{cases} \, ,
\eea
and
\bea
 \label{em:charges:4d:scalars}
 q_{0} & = - e^{2 \left( V - U \right)} \tilde{q}'^{0} \im \Pi
 \left[ 1 - \sum_{I = 1}^{3} \frac{\re z^I}{\left( \im z^I \right)^2}
 \left( \frac{\tilde{q}'^{I}}{\tilde{q}'^{0}} - \re z^I \right) \right] \\
 & - 2 p^0 \re \Pi + \sum_{\substack{ I < J \\ ( \neq K)}} \re z^I \re z^J p^K \, , \\
 q_I & = - e^{2 \left( V - U \right)} \tilde{q}'^{0} \frac{\im \Pi}{\left( \im z^I \right)^2}
 \left( \frac{\tilde{q}'^{I}}{\tilde{q}'^{0}} - \re z^I \right)
 + \frac{\re \Pi}{\re z^I} \bigg( p^0 - \sum_{J (\neq I)} \frac{p^J}{\re z^J} \bigg) \, ,
\eea
where we employed the following notation
\bea
 \im \Pi \equiv \im z^1 \im z^2 \im z^3 \, , \qquad \qquad
 \re \Pi \equiv \re z^1 \re z^2 \re z^3 \, .
\eea

\section{Generalities about the supersymmetric Casimir energy}
\label{AppC}

The partition function of an $\cN = 1$ supersymmetric gauge theory with a non-anomalous $\U(1)_R$ symmetry
on a Hopf surface $\cH_{p, q} \simeq S^3 \times S^1$, with a complex structure characterized by two parameters $p$, $q$, can be written as
\bea\label{Zindex}
 Z \left[ \cH_{p,q} \right] = e^{ - E_{\text{SUSY}}} I(p, q) \, .
\eea
Here, $I(p, q)$ is the superconformal index \cite{Kinney:2005ej,Romelsberger:2005eg} 
\bea I(p, q)= {\rm  Tr} (-1)^F  p^{  h_1 + r/2} q^{h_2+r/2} \, ,\eea
where $h_1$ and $h_2$ are the generators of rotation in orthogonal planes and $r$ is the superconformal R-symmetry.
$E_{\text{SUSY}}$ is the supersymmetric Casimir energy \cite{Assel:2014paa,Lorenzen:2014pna,Assel:2015nca},
\bea\label{Ecas}
 E_{\text{SUSY}} (b_1 , b_2) = \frac{4 \pi}{27} \frac{\left( |b_1| + |b_2| \right)^3}{|b_1| |b_2|} \left( 3 c - 2 a \right) - \frac{4 \pi}{3} \left( |b_1| + |b_2| \right) \left( c - a \right) \, ,
\eea 
where $p = e^{- 2 \pi |b_1|}$, $q = e^{- 2 \pi |b_2|}$, and $a$, $c$ are the central charges of the four-dimensional $\cN = 1$ theory.
We can extrapolate this result to include flavor symmetries by considering $a$ and $c$ as trial central charges,
depending on a set of chemical potentials $\hat\Delta_I$,
\bea\label{central charges}
 a(\hat \Delta_I) = \frac{9}{32}  {\rm  Tr} R(\hat \Delta_I)^3 - \frac{3}{32}{\rm  Tr} R(\hat \Delta_I) \, ,
 \qquad  c(\hat \Delta_I) = \frac{9}{32}  {\rm  Tr} R(\hat \Delta_I)^3 - \frac{5}{32}{\rm  Tr} R(\hat \Delta_I) \, ,
\eea
where $R$ is a choice of $\U(1)_R$ symmetry and the trace is over all fermions in the theory.
The supersymmetric Casimir energy can be also interpreted as the vacuum 
expectation value $\langle H_{\rm susy} \rangle$ of the Hamiltonian which generates time translations \cite{Assel:2015nca}.
It can also be obtained by integrating anomaly polynomials in six dimensions \cite{Bobev:2015kza}.
In particular, the supersymmetric Casimir energy for $\cN = 4$ super Yang-Mills (SYM) theory with $\SU(N)$ gauge group,
where $a=c$, reads\footnote{The R-symmetry 't Hooft anomalies for $\cN = 4$ SYM are given by
$\Tr R(\hat \Delta_I)=(N^2-1) [\sum_{I=1}^3 (\hat\Delta_I-1)+1]=0$ and
$\Tr R(\hat \Delta_I)^3 = (N^2-1)  [\sum_{I=1}^3 (\hat\Delta_I-1)^3+1] =3 (N^2-1) \hat \Delta_1 \hat \Delta_2 \hat \Delta_3$.
The $\hat\Delta_I$ are the R-symmetries of the three adjoint chiral multiplets of $\cN=4$ SYM and
satisfy (\ref{constraint2:N=4}).}
\bea
 \label{Casimir:SUSY2:N=4}
 E_{\text{SUSY}} = \frac{\pi}{8} \left( N^2 - 1 \right) \frac{\left( |b_1| + |b_2| \right)^3}{|b_1| |b_2|} \hat \Delta_1 \hat \Delta_2 \hat \Delta_3 \, ,
\eea
where $\hat \Delta_{1, 2, 3}$ are the chemical potentials for the Cartan generators of the R-symmetry, fulfilling the constraint
\bea
 \label{constraint2:N=4}
 \hat \Delta_1 + \hat \Delta_2 + \hat \Delta_3 = 2 \, . 
\eea
We can rewrite  Eq.\,\eqref{Casimir:SUSY2:N=4} in the notation used in the main text as 
\bea
  \label{modified:Casimir:SUSY2:N=4}
 E_{\text{SUSY}} = - i \pi  (N^2-1)  \frac{\Delta_1 \Delta_2 \Delta_3}{\omega_1 \omega_2} 
 \, ,
\eea
where we defined $\Delta_I =i  \left( |b_1| + |b_2| \right) \hat \Delta_I / 2$ and $\omega_i=-i |b_i|$. 
They satisfy the constraint
\bea
 \label{modified:constraint2:N=4}
 \Delta_1 + \Delta_2 + \Delta_3 +\omega_1+\omega_2 = 0 \, .
\eea
When extended to the complex plane $\Delta_I$ and $\omega_i$ are defined modulo 1. 

The constraints \eqref{modified:constraint2:N=4} and \eqref{constraint:N=4} are closely related to the absence of pure and mixed gauge anomalies. Let us briefly see why.
Consider again a generic $\cN=1$ supersymmetric gauge theory. The constraint \eqref{modified:constraint2:N=4} is modified to
\bea
 \label{modified:constraint3:N=4}
\sum_{I\in W_a} \Delta_I +\sum_{i=1}^2 \omega_i = 0 \, ,
\eea
where $\Delta_I$ is the chemical potential for the $I$-th field and the sum is restricted  to the fields entering the superpotential term $W_a$.
There is one constraint for each monomial $W_a$ in the superpotential of the theory.
The path integral on $S^3\times S^1$ localizes to a matrix model where one-loop determinants must be regularized.
$E_{\text{SUSY}}$  arises from the following regularization factors \cite{Assel:2015nca,Bobev:2015kza}
\bea \Psi(u)=  i \pi \frac{(\sum_{i=1}^2 \omega_i)^3}{24 \omega_1\omega_2}\left [ ( \hat u -1)^3 -\frac{\sum_{i=1}^2 \omega_i^2}{(\sum_{i=1}^2 \omega_i)^2} (\hat u-1)\right] \, ,\eea
where $\hat u =- 2 u/\sum_{i=1}^2\omega_i$ and $u$ is a chemical potential for gauge or flavor symmetries.
More precisely, denoting with $z$ the gauge variables, $E_{\text{SUSY}}$ gets an additive contribution $\Psi(z+\Delta_I)$ from each chiral multiplet and $-\Psi(z)$ for each vector multiplet. One can pull out regularization factors  from the matrix model only if they are independent of the gauge variables.
The constraint \eqref{modified:constraint3:N=4} implies that $\sum_{I\in W_a} \hat \Delta_I=2$, where again $\hat \Delta_I =- 2 \Delta_I/\sum_{i=1}^2\omega_i$.
Hence $\hat\Delta_I$ parameterizes a trial R-symmetry of the theory. One can see that, if all (pure and mixed) gauge anomalies cancel,  $E_{\text{SUSY}}$ is indeed independent of $z$ if the chemical potential satisfy \eqref{modified:constraint3:N=4} \cite{Bobev:2015kza}. The final result for 
$E_{\text{SUSY}}$ is then easily computed to be,
\bea\label{ES}  E_{\text{SUSY}} =  i \pi \frac{(\sum_{i=1}^2 \omega_i)^3}{24 \omega_1\omega_2}\left [ {\rm  Tr} R(\hat \Delta_I)^3 -\frac{\sum_{i=1}^2 \omega_i^2}{(\sum_{i=1}^2 \omega_i)^2} {\rm  Tr} R(\hat \Delta_I)\right] \, . \eea
Using \eqref{central charges} and $\omega_i=-i |b_i|$ we recover \eqref{Ecas}.

A similar quantity constructed from 
\bea \tilde\Psi(u)= i \pi \frac{(\sum_{i=1}^3 \omega_i)^3}{24 \omega_1\omega_2\omega_3}\left [ ( \hat u -1)^3 -\frac{\sum_{i=1}^3 \omega_i^2}{(\sum_{i=1}^3 \omega_i)^2} (\hat u-1)\right] \, ,\eea
 appears in the modular transformation of the integrand of the matrix model \cite{Brunner:2016nyk}.\footnote{This can be expressed in terms of Bernoulli polynomials as $\tilde\Psi(u)=  \frac{\pi i}{3} B_{3,3}(u; \omega_i)$  \cite{Brunner:2016nyk}.} Here the angular momentum fugacities are written as $p=e^{-2\pi i \omega_1/\omega_3}, q=e^{-2\pi i \omega_2/\omega_3}$,  the gauge fugacity as  $e^{2\pi u/\omega_3}$, and
$\hat u =- 2 u/\sum_{i=1}^3\omega_i$. In a theory with no gauge anomalies, the sum of $\tilde\Psi(z+\Delta_I)$ from each chiral multiplet and $-\tilde\Psi(z)$ for each vector multiplet is  independent of the
gauge variables $z$  and can be pulled out of the integral if the constraint
\bea
 \label{modified:constraint4:N=4}
\sum_{I\in W_a} \Delta_I +\sum_{i=1}^3 \omega_i = 0 \, ,
\eea
is fulfilled \cite{Spiridonov:2012ww,Brunner:2016nyk}. The sum then becomes
\bea
 \label{FS}
 \varphi=  i \pi \frac{(\sum_{i=1}^3 \omega_i)^3}{24 \omega_1\omega_2\omega_3}
 \left [ {\rm  Tr} R(\hat \Delta_I)^3 -
 \frac{\sum_{i=1}^3 \omega_i^2}{(\sum_{i=1}^3 \omega_i)^2} {\rm  Tr} R(\hat \Delta_I)\right] \, .
\eea
This observation has been used in  \cite{Brunner:2016nyk} to write the index as
\bea
 \label{FS2}
 I(p,q,\Delta_I) = e^{\varphi} I^{{\rm mod}} (\omega_i,\Delta_I) \, ,
\eea
where $I^{{\rm mod}}$ is a modified matrix model depending on modified elliptic gamma functions. 
As noticed in \cite{Brunner:2016nyk}, the supersymmetric Casimir energy can be extracted from $\varphi$ in the low temperature limit. 
Indeed, in the limit $\omega_3 = 1/\beta \to 0$, $\varphi$  becomes exactly $\beta E_{\text{SUSY}}$.
It is interesting to notice that, for $\cN = 4$ SYM,  by setting one of the $\omega_i$ equal to $-1$,   $\varphi$ reduces to our quantity (\ref{modified:Casimir:SUSY:N=4}) and \eqref{modified:constraint4:N=4}  to the constraint \eqref{constraint:N=4}. 
This is an observation to explore further. In particular, it would be interesting to understand the physical meaning of $I^{{\rm mod}}$ and  the decomposition
(\ref{FS2}).

\bibliographystyle{ytphys}

\bibliography{AdS5_BH_Casimir}{}

\end{document}